\documentclass[11pt,twoside]{article}

\usepackage{asp2006}
\usepackage{epsf}
\usepackage{psfig}
\usepackage{lscape}
\usepackage{amsmath}
\usepackage{amssymb}
\usepackage{natbib}
\usepackage{epsfig}

\def\Fe{\mathrm{Fe}}
\def\feh{[\Fe/\mathrm{H}]}
\def\Nafe{[\mathrm{Na}/\mathrm{Fe}]}
\def\Ne22{^{22}\mathrm{Ne}}
\def\Na23{^{23}\mathrm{Na}}
\def\ne22pg{\Ne22(p,\gamma ) \Na23}
\def\msun{\mathrm{M}_{\odot}}

\def\dex{\mathrm{dex}}

\begin{document}
\title{Galactic Sodium from AGB Stars}   
\author{Robert G. Izzard$^1$, Brad K. Gibson$^2$ and Richard J. Stancliffe$^3$}   
\affil{
1 Sterrenkundig Instituut, University of Utrecht, The Netherlands\\
2 University of Central Lancashire, UK\\
3 Institute of Astronomy, University of Cambridge, UK\\
}    

\begin{abstract} 
Galactic chemical evolution (GCE) models which include sodium from type~II supernovae (SNe) alone underestimate the abundance of sodium in the interstellar medium by a factor of two to three over about $3\dex$ in metallicity and predict a flat behaviour in the evolution of $\Nafe$ at super-solar metallicities. Conversely, recent observations of stars with $\feh \sim +0.4$ suggest that $\Nafe$ increases at high metallicity. We have combined stellar evolution models of asymptotic giant branch (AGB) and Wolf-Rayet (WR) stars with the latest SN yields in an attempt to resolve these problems \dots and have created many more.
\end{abstract}



\vspace{-2mm}

\section{Introduction}
Recent observations of sodium in stars with metallicities up to $\feh=+0.4$ suggest secondary sodium production at high $Z$ \citep{gilli}. Is it a chemical evolution effect and if so, what is the source of the sodium? Canonical GCE models (e.g. \citealp{timmes}) underproduce sodium by $\sim 0.2\,\dex$, but only include massive star yields.  There is observational scatter, up to $0.5 \,\dex$, which is much larger than observational errors. How can we make interstellar gas with $\feh\sim+0.4$?
We use stellar models in combination with a dual-infall GCE code to try to reproduce the observed sodium abundance trends.

\vspace{-2mm}

\section{Stellar Models}

Our models with $0.1\leq M\leq 8 M_\odot$ include 1st, 2nd and 3rd dredge-ups, hot-bottom burning (HBB) and mass-loss \citep{izzard06}. AGB stars with $4\lesssim M/\msun \lesssim 8$  make sodium by HBB via the $\ne22pg$ reaction. The rate is uncertain so we vary it within experimental limits (up to $\times 2000$). For massive stars ($8\leq M/\mathrm{M}_\odot \leq 80$) we use the STARS code with a full nucleosynthetic network. Mass loss during the WR phase exposes layers which have converted $\Ne22$ to $\Na23$ by hydrogen burning. $\Na23$ is made during carbon burning in massive stars and is ejected in the SN explosion at the end of the star's life. We use the $\Na23$ yields of \citet{chieffi04}, which are a strong function of mass and metallicity. Figure 1 (left) shows the time- and mass-integrated ejecta from our stellar populations as a function of metallicity $Z$.  We use the IMF of \citet{KTG93}.


\section{GCE Models}
We have implemented our stellar models and SNeIa into the \cite{chiappini} dual-infall homogeneous GCE model adopting the instantaneous mixing approximation without radial flows.We tried five sets of physical parameters: \textbf{1.}~The standard dual-infall model with our standard stellar models, \textbf{2.}~As 1 with SN yields of sodium $\times {1}/{2}$, \textbf{3.}~As 2 with $\ne22pg$ rate $\times 10$, \textbf{4.}~As 2 with $\ne22pg$ rate $\times 2000$ and \textbf{5.}~As 2 with infall from a polluted instead of primordial gas.

We find that:
\textbf{1.}~Reduction of sodium yield is necessary at solar metallicity, but not for $\feh <-1$.
\textbf{2.}~We find a local minimum in $\Nafe$ which qualitatively matches the observations, albeit at too low $\feh$. 
\textbf{3.}~A $2000\times$ increase in the $\ne22pg$ rate gives too much $\Na23$, but $\times 10$ is compatible with the observations.
\textbf{4.}~It is hard to make gas with $[\mathrm{Fe}/\mathrm{H}]>0$ unless we include feedback or thaumaturgy.
\textbf{5.}~We cannot reproduce the scatter in $\Nafe$. 

\vspace{-3mm}\hspace{-16mm}\begin{tabular}{cc}
\includegraphics[scale=0.28,angle=270]{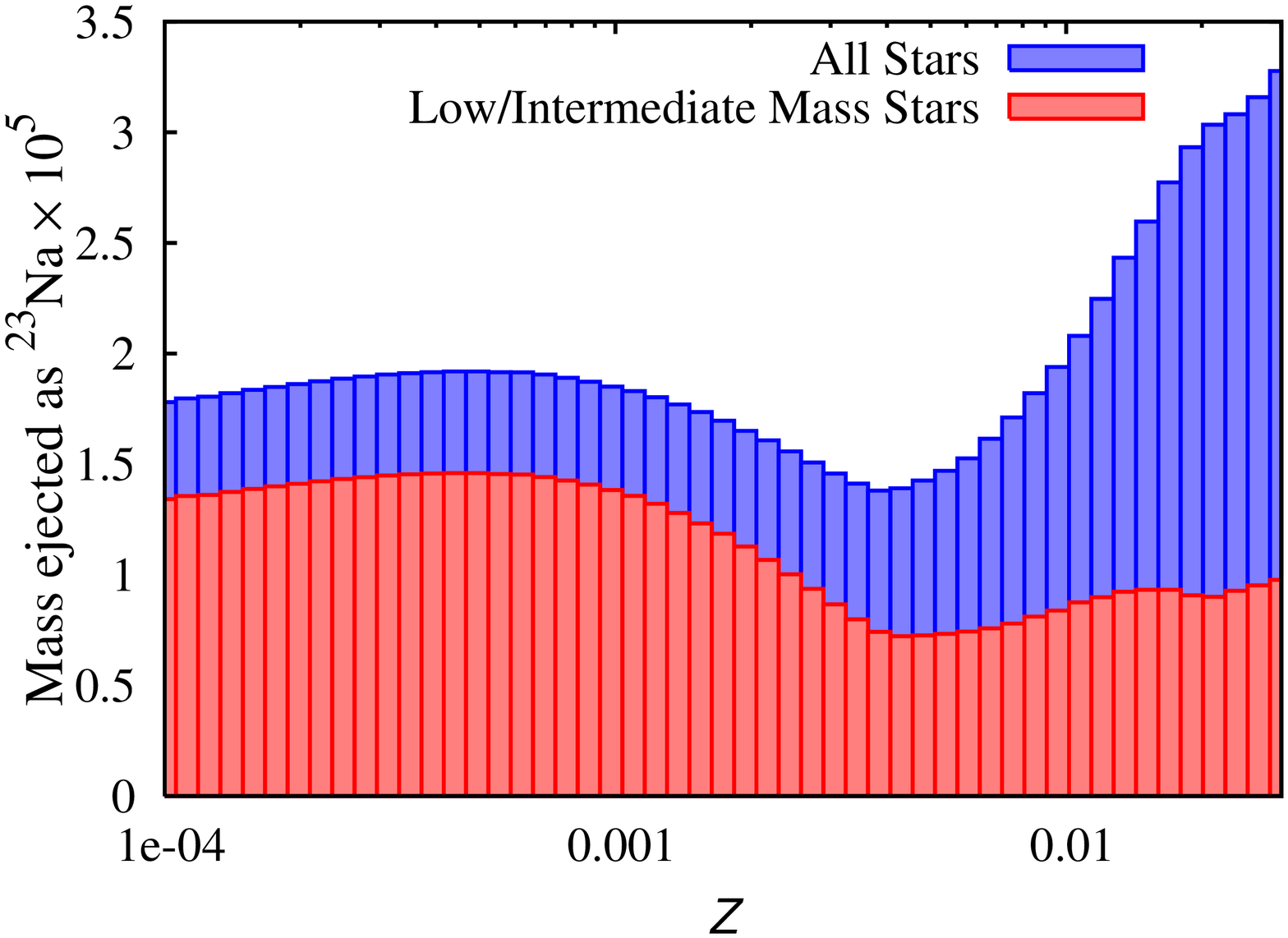}
&
\includegraphics[scale=0.28,angle=270]{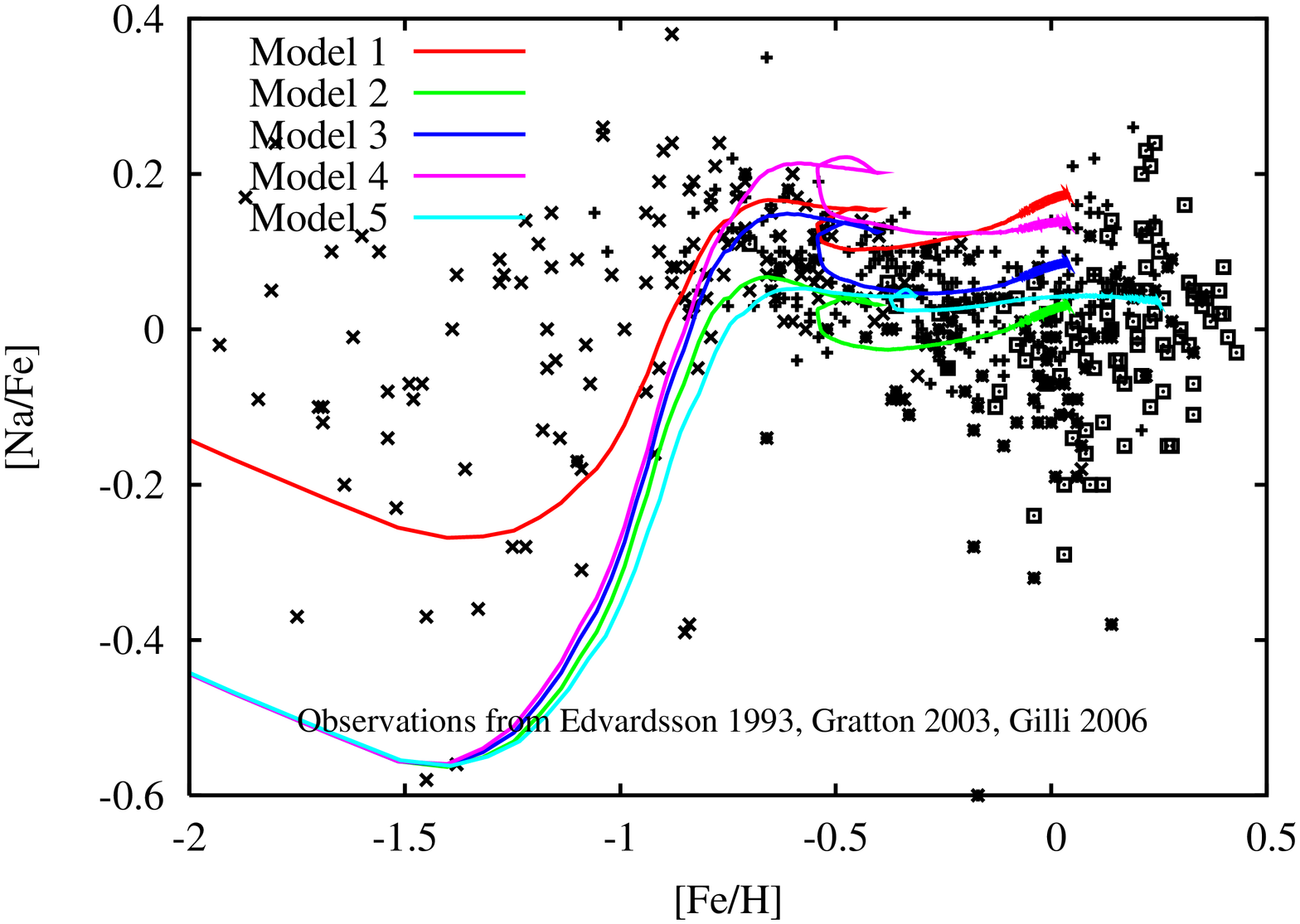}
\\
\end{tabular}
\begin{footnotesize}
Figure 1: \begin{minipage}[t][1\totalheight]{1\columnwidth}Left: Ejecta of sodium as a function of $Z$ from intermediate mass and all stars.\\ Right: $\Nafe$ vs time in our GCE simulations. \end{minipage}
\end{footnotesize}


\vspace{-5mm}

\section{Conclusions}

\vspace{-1mm}GCE models of sodium should include the contribution from AGB stars undergoing HBB, but this leads to overproduction of sodium in the disk. Our GCE models qualitatively match the observed $\Nafe$ vs $\feh$, fail to reproduce the observed scatter in $\Nafe$ and never make gas with $\feh\sim +0.4$. The hint of an increase in $\Nafe$ at high $\feh$ may be due to secondary $\mathrm{Na}$ from type~II SNe, not AGB stars.

\acknowledgements 
RGI is supported by the NWO, BKG is supported by PPARC, RJS thanks Churchill College for his fellowship.

\vspace{-4mm}
\bibliographystyle{plain}

\end{document}